\documentclass[prl,epsf]{revtex4}
\usepackage{graphicx}
\usepackage{epsfig}
\usepackage{amsfonts}
\usepackage{amsmath}
\usepackage{amssymb}
\begin{document}
\title{Planar spin exchange in LiNiO$_2$ }
\author{A.-M. Dar\'e, and R. Hayn}
\affiliation{
Laboratoire Mat{\'e}riaux et Micro{\'e}lectronique de
Provence UMR 6137, B\^at. IRPHE, Technop\^ole de Ch\^ateau Gombert,
49 rue Joliot Curie BP 146,
13384 Marseille Cedex 13, France}
\date{\today}

\begin{abstract}
We study the planar spin exchange couplings in LiNiO$_2$ using 
a perturbative approach.
We show that the inclusion 
of the trigonal crystal field splitting at the Oxygen sites leads to the 
appearance of antiferromagnetic exchange integrals in deviation from the 
Goodenough-Kanamori-Anderson rules for this 90 degree bond. That gives 
a microscopic foundation for the recently observed coexistence of 
ferromagnetic and antiferromagnetic couplings in the orbitally-frustrated 
state of LiNiO$_2$ (F.\ Reynaud et al, Phys.\ Rev.\ Lett.\ {\bf 86}, 
3638).
\end{abstract}
\maketitle

The compound LiNiO$_2$ which was first synthesized in 1958 
\cite{goodenough58} is known to be a good ionic conductor and therefore 
suitable as a material for rechargeable batteries. Despite its wide use as 
electrode material, its electronic and magnetic structures 
are not yet completely understood.
Especially intriguing is the absence of any kind of long range magnetic or
orbital order \cite{yamaura96,reynaud01} at low temperature even in the 
purest samples synthesized up to now. That is especially remarkable since the 
isostructural compound NaNiO$_2$ shows orbital order and a collective 
Jahn-Teller transition at $T_o=480$ K from the trigonal high-temperature 
phase to the monoclinic low-temperature one, followed by an antiferromagnetic
(AFM) order of ferromagnetic (FM) planes at the N\'eel temperature of $T_N=20$ K 
\cite{bongers66,chappel00}. 
To explain the strange behavior of LiNiO$_2$ the proposal of an orbital 
liquid was pursued in terms of the SU(4) model \cite{feiner97,li98,bossche00}. 
There, a symmetry between orbital and spin degrees of freedom is assumed with 
equal amplitudes for the corresponding coupling terms. 
In reality, however, the energy scale for orbital interactions is one 
order of magnitude larger than those for spin exchange interactions 
as shown by experimental \cite{reynaud01} and theoretical studies 
\cite{mostovoy02}. This is also indicated by the difference of $T_o$ 
and $T_N$ in NaNiO$_2$. More in details, the magnetic 
susceptibility of LiNiO$_2$ shows a transition at 
$T_{of}=400$ K towards an orbitally frustrated state at low 
temperature. Given an orbital disorder that is frozen in, the magnetic 
properties at low temperature can be phenomenologically explained 
assuming FM exchange couplings between neighbouring orbitals of 
different kind ($J_{do}=(-6.2 \pm 2)$ meV) and 
AFM couplings between identical ones 
($J_{so}=(6.9 \pm 2)$ meV) \cite{reynaud01}. 
However, that contradicts seemingly the known 
Goodenough-Kanamori-Anderson (GKA) rules 
\cite{GKA} which 
allow only FM couplings for the 90 degree Ni-O-Ni bond in 
LiNiO$_2$ \cite{mostovoy02}.

The trigonal (rhombohedral) crystal structure $R\bar{3}m$ of LiNiO$_2$ 
can be understood by starting from the cubic situation with oxygen and Ni/Li 
on the sites of the cubes, and with the cubic axes $\tilde{x}$, 
$\tilde{y}$, and $\tilde{z}$.
Perpendicular to the cube diagonal 
$z=\tilde{x}+\tilde{y}+\tilde{z}$ in fig.\ 1 one finds 
alternating planes of Li, Ni, and O. The electronically active NiO$_2$ 
layer (see fig.\ 1) contains 
a triangular lattice of  
the magnetic Ni ions. The O$^{2-}$ ions have 
a completely filled 2$p$ shell, whereas the Ni$^{3+}$ ion 
is in the low-spin 
electronic configuration ($t_{2g}^6$ $e_g^1$) with spin 1/2
\cite{reynaud01}.

The trigonal 
distortion changes the bond angle from the ideal value of $90^{\circ}$ to 
$94^{\circ}$, but the six neighbouring oxygens of Ni stay equivalent. Also 
the Ni $e_g$ orbitals remain degenerate, even in the trigonal crystal field.  
We now assume that we have some kind
of orbital disorder that is frozen in for the triangular Ni lattice. The 
present knowledge does not conclusively predict in which way the 
orbital degeneracy is locally broken. 
Therefore, we have investigated two situations. The first case deals 
with orbitals that are oriented along the cubic axes:
$d_{\tilde{x}^2-\tilde{y}^2}$ (0) and $d_{3 \tilde{z}^2-r^2}$ (1). 
The second case chooses two complex combinations of the above orbitals 
that were predicted in Refs.\ \cite{brink01,khomskii,maezono00} and which have the 
advantage that they preserve the trigonal symmetry. 
We will now show that neglecting the crystal field splitting at the 
Oxygen sites one finds exclusively FM nearest 
neighbour exchange couplings. Due to the alternating stacking of Li and 
NiO$_2$ layers, however, one expects a considerable trigonal crystal 
field splitting of 
the O 2$p$ orbitals (into one doublet and a singlet). That leads to the 
possibility of AFM exchange integrals in the Ni plane. 

\begin{figure}
\includegraphics[width=10cm,angle=0]{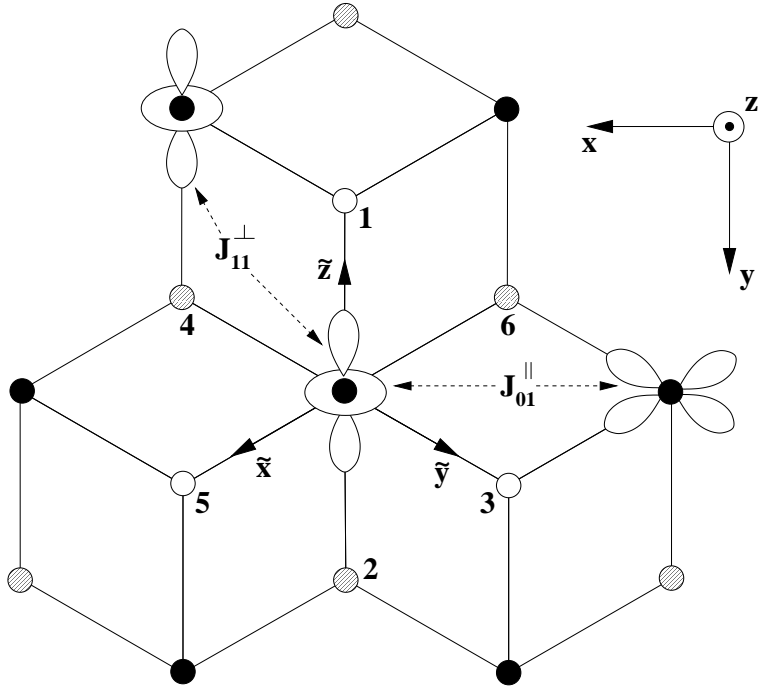}
\caption{
Part of the NiO$_2$ layer consisting of the Ni plane (filled circles) and two 
O planes above (open circles) and below (hatched circles) the Ni one. The two 
coordinate systems used in the calculation are oriented along the cubic 
($\tilde{x}$, $\tilde{y}$, $\tilde{z}$) or trigonal ($x$, $y$, $z$) axes. 
Also shown are three orbitals oriented along the cubic $\tilde{z}$-axis with the 
corresponding notation of exchange integrals.}
\label{fig1.lbl}
\end{figure}

In the actual calculations we take
the Hamiltonian in hole notation. The local part 
\begin{equation}
H^0=H^0_d +H^0_p
\label{general} \end{equation}
contains the Nickel contribution, which for one atom (omitting the site index) 
is given by
\begin{equation}
H^0_d=e_d \sum_{m} \hat{n}_{dm}+U_d \sum_{m} \hat{n}_{dm\uparrow} 
\hat{n}_{dm\downarrow}
+{U_{di} \over 2}  \sum_{m} \hat{n}_{dm}
\hat{n}_{d \bar{m}}
- {J_{Hd} \over 2} \sum_m ( 2\vec{S}_{dm} \vec{S}_{d\bar{m}} + {1\over 2}
\hat{n}_{dm}\hat{n}_{d\bar{m}} )
\label{hnickel}
\end{equation}
where $m$ and $\bar{m}$ stand for the two orthogonal $e_g$ orbitals.
$U_d$ and $U_{di}$ are respectively the inter and intra-orbital 
Coulomb repulsions,
which are equal when the Hund's intra-atomic exchange $J_{Hd}$ is neglected. 
For an Oxygen atom the local Hamiltonian is
\begin{equation}
H^0_p=\sum_{n} \epsilon_{n} \hat{n}_{pn}
+U_p \sum_{n} \hat{n}_{pn\uparrow}
\hat{n}_{pn\downarrow}
+ {U_{pi} \over 2}\sum_{n,\bar{n} }\hat{n}_{pn}
\hat{n}_{p\bar{n}}
- {J_{Hp} \over 2} \sum_{n,\bar{n}} (2 \vec{S}_{pn} \vec{S}_{p\bar{n}} 
+{1\over 2} \hat{n}_{pn}\hat{n}_{p\bar{n}} ) \; , 
\label{hoxygen}
\end{equation}
where $n$ and $\bar{n}$ are two of the three different $2p$ orbitals.
$U_p$ and $U_{pi}$ are the intra and inter 
orbital Hubbard-like repulsions, respectively and 
$J_{Hp}$ is the Hund's exchange integral. 
It is well known for $p$ as well as for $e_g$ orbitals,
using the definitions of $U_i$, $U$ and $J_H$ as Coulomb's integrals,
that one finds the general relation
$U_{i}=U-2J_{H}$. However, the rotational symmetry at each site
for degenerate orbitals demands an 
Hamiltonian of a more general form than the one used in eqs. (\ref{hnickel}, \ref{hoxygen})
\cite{fazekas,cyrot}. One can keep the truncated parts as in eqs. 
(\ref{hnickel}, \ref{hoxygen}), but then one has to use instead the relation
$U_{i}=U-J_{H}$ to restore, in the case of degenerate orbitals,
the invariance of the Hamiltonian under
arbitrary rotation of the basis at each site.

The hopping term between Nickel and Oxygen involves the 6 O of the NiO$_6$
octahedra.
Indexing the Ni site by $x$, the Ni orbital by $m$, the O site by $i$, 
and the O orbital by $n$,
the hybridization part of the Hamiltonian reads
\begin{equation}
H_{dp}=-\sum_{x,m,\sigma }\sum_{i,n} ( t_{xi}^{mn} 
p^{\dagger}_{in\sigma} d_{xm\sigma} + t_{ix}^{nm} d^{\dagger}_{xm\sigma} p_{in\sigma}) \ ,\; 
\label{hopping} 
\end{equation}
where $t_{ix}^{nm}=(t_{xi}^{mn})^{\ast}$. The hopping parameters are detailed below.  
Direct hopping between Ni is neglected.
We evaluate the spin exchange between two Ni at the fourth-order 
in perturbation theory 
in the hopping parameters, 
looking for an effective Hamiltonian of the form 
\begin{equation}
H^{eff}= J_{mm'} {\vec S}_{x} {\vec S}_{x'} \ ,
\end{equation}
$m$ and $m'$ are the orbital indices of the unpaired holes, $x$ and $x'$ the Nickel site labels. 
In hole representation, the 2$p$ levels are empty, while there are 
three fermions per site in the $e_g$ levels. If we neglect 
Hund's local intra-orbital exchange $J_{Hd}$ at Ni, we find in 
fourth-order perturbation theory the same processes
that for the case of one fermion at each Ni site. 
To perform perturbation theory, we make no assumption about the 
orbital order, looking at all the possibilities for
$m$ and $m'$, 
but  we suppose that the orbital degree of freedom is frozen in 
the spin exchange process: in the initial and final 
states the unpaired fermion on Ni is on the same $m$ orbital.
Neglecting $J_{H d}$, the result for the spin exchange is:
\begin{eqnarray}
J_{mm'} &=&
{\frac 4  {U_d}} {\Biggl |}\sum_{n} { t^{ \ m'}_{nx'} t^{m}_{xn}  
\over \tilde{\epsilon}_{n} } {\Biggr |}^2 \nonumber \\
&+& 2 \sum_{n} \sum_{n' \neq \bar{n}}
t^{\ m'}_{nx'}  t^{\ m}_{n'x}  t^{m'}_{x'n'}  t^m_{xn}
{\Biggl (} {1 \over \tilde{\epsilon}_n} 
+{1 \over \tilde{\epsilon}_{n'}} {\Biggr )}^2
{ 1 \over \tilde{\epsilon}_n +\tilde{\epsilon}_{n'} 
+ U_p \delta_{nn'}}\nonumber \\
&+&  \sum_{n} \sum_{\bar{n}}  t^{\ m'}_{nx'}  
t^{\ m}_{\bar{n}x}  t^{m'}_{x' \bar{n}} t^m_{xn}
{\Biggl (} {1 \over \tilde{\epsilon}_n} 
+{1 \over \tilde{\epsilon}_{\bar{n}}} {\Biggr )}^2  
{\Bigl(} {1 \over  \tilde{\epsilon}_n 
+\tilde{\epsilon}_{\bar{n}} +U_{pi}-J_{Hp}}
+{1 \over \tilde{\epsilon}_n +\tilde{\epsilon}_{\bar{n}}+ U_{pi}+J_{Hp}}
{\Bigr )}\nonumber \\
&-&  \sum_{n} \sum_{\bar{n}} {\Bigl |}  t^{m'}_{x'\bar{n}} t^m_{xn}  
{\Bigr |}^2
{\Biggl (} {1 \over \epsilon_n} +{1 \over \epsilon_{\bar{n}}} {\Biggr )}^2 
{\Bigl(} {1 \over \tilde{\epsilon}_n +\tilde{\epsilon}_{\bar{n}} 
+U_{pi}-J_{Hp}}
-{1 \over \tilde{\epsilon}_n +\tilde{\epsilon}_{\bar{n}}+ U_{pi}+J_{Hp}}
{\Bigr )}  \ ,
\label{jmm'}
\end{eqnarray}
where the energy needed to transfer a hole from Nickel to Oxygen is 
$\tilde{\epsilon}_n= {\epsilon}_n-e_d-2 U_d$.
For a Ni pair, sharing two Oxygen, 6 $2p$ levels labelled by $n$ are 
involved in the intermediate states. 
The sum on $n$ enables to drop the Oxygen site 
index used in eq.~(\ref{hoxygen}),
$n$ and ${\bar{n}}$ refer to two different orbitals of the same O site. The crystal 
field splitting at the Oxygen sites 
is taken into account through $\tilde{\epsilon}_n= \Delta -\delta$ for 
two levels per Oxygen,
and $\tilde{\epsilon}_n= \Delta +\delta$ for the third one.
The first term of eq.~(\ref{jmm'}) implies a hopping of a hole from 
the first to the second Nickel;
in the second process, the two holes jump to the same $2p$ orbital, 
or to two different Oxygen sites.
The last two terms represent a process where the two holes meet on the same 
Oxygen, but on 
different orbitals, then the spin flip can act (fourth) or not (third). 
Including $J_{H d}$ one finds many additional terms involving spin-flip 
at the Ni sites. 
However these terms are typically smaller than the previous ones by 
factors of ${J_{H d} \over U_d}$ or ${J_{H d} \over \tilde{\epsilon}_n}$. 
We have explicitely checked that there is no qualitative change 
in the results when we add these terms  
for $J_{H d} \sim U_d /10$.

Considering the exchange integrals for the usual $e_g$ orbitals: 
{$d_{\tilde{x}^2-\tilde{y}^2}$ 
$(0)$ and $d_{3{\tilde{z}}^2-r^2}$ $(1)$  
it can be seen from fig.\ 1 that not only the two Ni orbitals need to be 
specified, but also the
particular Ni pair: this leads to 6 different spin exchange terms
$ J_{00}^{\perp}, J_{00}^{\parallel}, 
J_{01}^{\perp}, J_{01}^{\parallel},J_{11}^{\perp}, J_{11}^{\parallel}$. 
The two degenerate $2p$ orbitals are labelled $p_x$ and $p_y$, 
the last one $p_z$. 
In terms of the cubic $p_{\tilde{x}}$, $p_{\tilde{y}}$, and $p_{\tilde{z}}$ 
orbitals they read
\begin{subequations}
\label{pxyz}
\begin{eqnarray}
p_x &=& {1 \over \sqrt{2}} 
( p_{\tilde{x}}-p_{\tilde{y}}), \label{equationa} \\ 
p_y &=& {1 \over \sqrt{6}} ( p_{\tilde{x}}+p_{\tilde{y}}- 2 p_{\tilde{z}}), 
\label{equationb} \\ 
p_z &=& {1 \over \sqrt{3}} ( p_{\tilde{x}}+p_{\tilde{y}}+p_{\tilde{z}}) 
\label{equationc} 
\end{eqnarray}
\end{subequations}
The hopping parameters between the Nickel orbitals $(0)$ and $(1)$ and 
these Oxygen orbitals, depending 
on the O atom involved, can be obtained by linear combinations of 
the hopping integrals between the Ni 
orbitals and $(p_{\tilde{x}},p_{\tilde{y}},p_{\tilde{z}})$ which are 
specified in table I. 
An arabic index enables to distinguish between all the 6 Oxygen of
the NiO$_6$ octahedra, in 
correspondence with fig.\ 1.

\begin{table}
\caption{\label{hoppingparameters} Hopping integrals for 
$d_{\tilde{x}^2-\tilde{y}^2}$  and $d_{3{\tilde{z}}^2-r^2}$}
\begin{tabular}{c||c|c|c||c|c|c}
\hline
\multicolumn{1}{c||}{O index} &
\multicolumn{3}{c||}{$d_{\tilde{x}^2-\tilde{y}^2}$ } &
\multicolumn{3}{c}{$d_{3{\tilde{z}}^2-r^2}$} \\
\hline
\multicolumn{1}{c||}{} &
\multicolumn{1}{c|}{$ p_{\tilde{x}}$} &
\multicolumn{1}{c|}{$ p_{\tilde{y}}$} &
\multicolumn{1}{c||}{$ p_{\tilde{z}}$} &
\multicolumn{1}{c|}{$ p_{\tilde{x}}$} &
\multicolumn{1}{c|}{$ p_{\tilde{y}}$} &
\multicolumn{1}{c}{$ p_{\tilde{z}}$} \\
\hline
$0_1$ & 0 & 0 & 0 & 0 & 0 & ${2 \over \sqrt{3} } t_0$\\
$0_2$ & 0 & 0 & 0 & 0 & 0 & $-{2 \over \sqrt{3} } t_0$\\
$0_3$ & 0 & $-t_0$ & 0 & 0 & $-{1 \over \sqrt{3} } t_0$ & 0\\
$0_4$ & 0 & $t_0$ & 0 & 0 & ${1 \over \sqrt{3} } t_0$ & 0\\
$0_5$ & $t_0$ & 0 & 0 & $-{1 \over \sqrt{3} } t_0$ & 0 & 0\\
$0_6$ & $-t_0$ & 0 & 0 & ${1 \over \sqrt{3} } t_0$ & 0 & 0\\
\hline
\end{tabular}
\end{table}

The parameters $U_d=9.5$ eV, $U_p=4.6$ eV, and $t_0=1.3$ eV were taken from 
Ref.\ \cite{dagotto96}. The charge transfer energy $\Delta=4$ eV was roughly 
estimated from the difference of the centers of gravity of the 
3$d$ and 2$p$ levels in a bandstructure calculation \cite{mertz02}, which 
allows also to estimate the crystal field splitting $\delta$ from the 
bandwidth of the upper 3$d$ $e_g$ band. Namely, for $\delta=0$ and 
only nearest neighbour hopping in the tight-binding Hamiltonian (eq.~(\ref{hopping})), 
the bandwidth would be zero. We obtained 
a value for $\delta$ of about 1 eV, but we keep $\delta$ here
as a parameter. 
The result for the different spin exchange couplings as a function of the crystal field 
splitting $\delta$ is presented in fig.\ 2.
For $\delta=0$, the only contribution to exchange is from $J_{Hp}$
($J_{Hp}=U_p/10$). We used $U_{pi}=U_p-J_{Hp}$ (see above).
A splitting of $|\delta| \stackrel{>}{\approx} 0.5$ eV 
can change the ferromagnetic character into a 
antiferromagnetic one for the exchange integrals $J_{00}^{\parallel}$, 
$J_{11}^{\perp}$, and  $J_{11}^{\parallel}$. (The exchange coupling 
$J_{00}^{\perp}$ is identical to zero in the present approximation.) 
The exchange terms between different orbitals $J_{01}^{\perp}$ and  
$J_{01}^{\parallel}$ stay ferromagnetic (see Ref.\ \cite{reynaud01}).

It seems as if an antiferromagnetic exchange integral between equal orbitals 
is in contradiction with the existence of ferromagnetic planes in NaNiO$_2$.
One should keep in mind however that the low temperature phase of NaNiO$_2$ 
is not trigonal but has a large monoclinic distortion which leads to a different
crystal field splitting at Oxygen. One could speculate that the monoclinic 
distortion in this coumpound can be described in our calculations by a reduced effective parameter 
$\delta_{eff}$ shifting all the exchange integrals into the ferrromagnetic region.

\begin{figure}
\includegraphics[width=10cm,angle=-90]{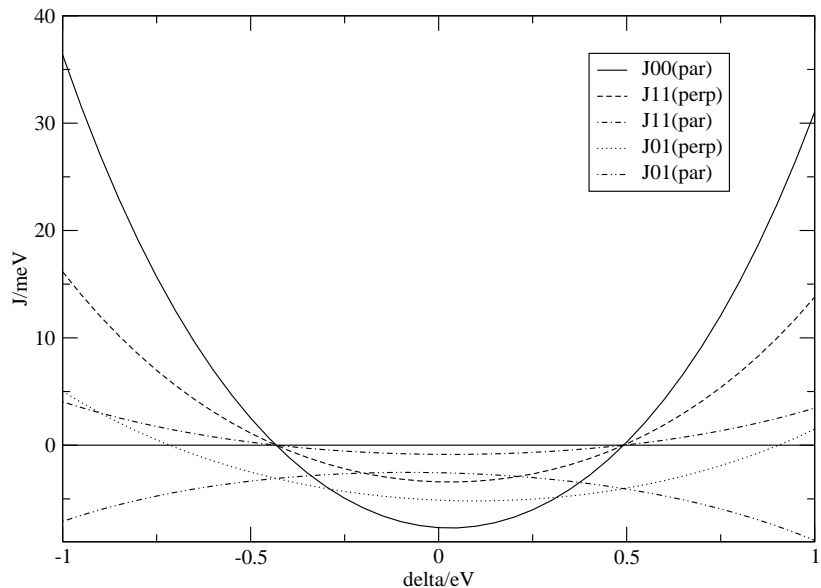}
\caption{Exchange couplings between orbitals oriented along one of the cubic axes}
\label{fig2.lbl}
\end{figure}

At present there is no experimental indication for macroscopic Jahn-Teller
distortions of the trigonal crystal structure of LiNiO$_2$ even at very low 
temperature.
Therefore it is tempting to search for an orbitally-frustrated state in terms of 
orbitals that obey the trigonal crystal symmetry. For that purpose, noting  
that the $e_g$ stay degenerate, we can use 
instead of $d_{\tilde{x}^2-\tilde{y}^2}$ and 
$d_{3{\tilde{z}}^2-r^2}$ any
linear and orthogonal combination of these two.
The particular choice
\begin{subequations}
\label{ronfleur}
\begin{eqnarray}
d_{\alpha} &=& {1 \over \sqrt{2}} ( d_{\tilde{x}^2-\tilde{y}^2} 
+i d_{3\tilde{z}^2-r^2} ),  \label{equationap} \\ 
d_{\beta} &=& {1 \over \sqrt{2}} ( d_{\tilde{x}^2-\tilde{y}^2} 
-i d_{3\tilde{z}^2-r^2} )\label{equationbp} 
\end{eqnarray}
\end{subequations}
corresponds to orbitals which satisfy the trigonal symmetry: under a 
rotation of ${2 \pi \over 3}$ 
around the axis perpendicular to the Ni plane (fig.\ 1), these 
states are just changed by a phase factor.
These complex orbitals were previoulsy proposed for 
manganites \cite{brink01,khomskii,maezono00}. It cannot be a priori excluded
that also in LiNiO$_2$ one of these complex orbitals is 
preferred at each Ni site. 

In terms of these new orbitals the calculations get simpler: 
there are only 
two different exchange integrals:
$J_s=J_{\alpha \alpha}= J_{\beta \beta}$ and 
$J_d=J_{\alpha \beta}$, and there is no need to consider 
different Ni pairs, due to the trigonal symmetry.
However it is not possible to establish a simple link between these two 
new exchange terms, and the 6 previously studied: this is
due to freezing the orbital degree of freedom in the calculation of the 
spin exchange. 
Using the transformation (\ref{ronfleur}) and the table (\ref{hoppingparameters})
one can easily specify the hopping for complex orbitals to be inserted in 
eq. (\ref{jmm'}).

The results are presented in fig.\ 3. At the same 
critical $\delta$- value as in fig.\ 2, the sign of $J_d$ 
changes to an AFM coupling for large $\delta$. But now, the 
effect is reversed in comparison to fig.\ 2 or Ref.\ \cite{reynaud01}, 
it is the exchange between {\em different} orbitals that becomes 
antiferromagnetic.

\begin{figure}
\includegraphics[width=10cm,angle=-90]{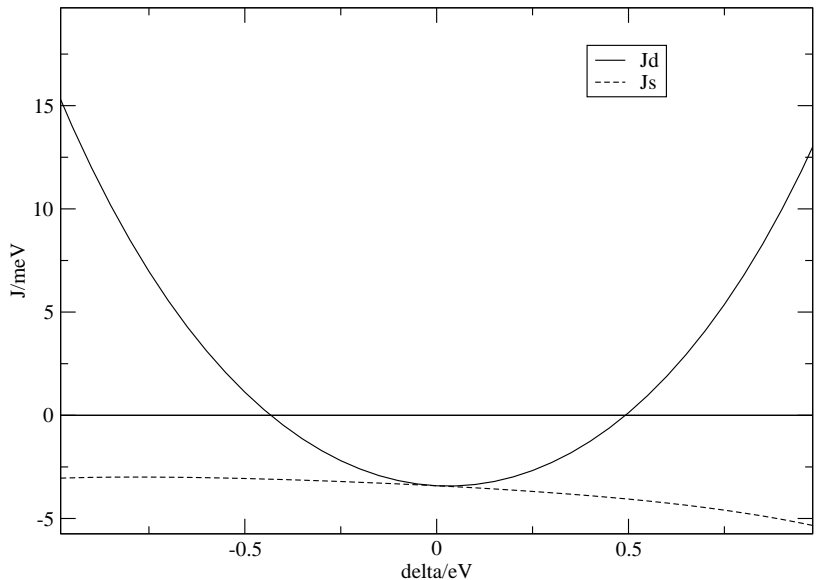}

\caption{Exchange couplings between orbitals oriented along the trigonal axis.
}
\label{fig3.lbl}
\end{figure}

In the basis of complex orbitals, we can also consider the pseudospin
exchange.
Due to pseudospin anisotropy, there are many terms in the effective 
Hamiltonian involving the pseudospin operators. 
These terms are usually of the order of 
$ U_p t_0^4 /(\Delta^3 (2 \Delta +U_p))$ for $\delta=0$. This confirms that they
are larger than the spin exchange terms characterized by $J_{Hp}$.
We have considered more in details the $z$ component of the pseudospin exchange 
which corresponds to a term $J^T_{z} T_i^zT^z_{j}$ for a pair of Ni sites labelled by $i$ and $j$,
with $T^z =+(-){1 \over 2}$  depending of the orbital involved $d_{\alpha}$ ($d_{\beta}$).
We suppose that the spin degree of freedom is frozen at each site.
Somewhat surprisingly when the Hund's exchange term for Oxygen, as well as as the splitting of the 
$2p$ levels are neglected, 
we obtain that $J^T_{z}=0$, indepently of the spins.
That shows the highly anisotropic character of the pseudospin exchange. 
But one should note that this result is not in contradiction with
the results from Mostovoy and Khomskii \cite{mostovoy02} 
(who found a positive value for the orbital exchange) since 
they analysed the usual  
$d_{\tilde{x}^2-\tilde{y}^2}$ and $d_{3\tilde{z}^2-r^2}$ orbitals  which 
are eigenstates of the $T^x$ pseudospin operators in the present basis.
The pseudospin-rotational symmetry being not required as for spin, 
the result $J^T_{z}=0$ does not rule out 
a pseudospin exchange along the others directions.
The evaluation of  other pseudospin-pseudospin or pseudospin-spin 
coupling terms will be detailed in a forthcoming paper using an alternative
method with a direct diagonalization of the hopping part of the 
Hamiltonian which enables to evaluate the different terms in a more 
direct way.

In conclusion, we have shown that a proper inclusion of the trigonal 
crystal field splitting at the Oxygen sites of LiNiO$_2$ leads to a 
coexistence of AFM and FM couplings. Orientating the orbitals along the 
cubic axes we find AFM couplings between equal orbitals and FM couplings 
between different ones, which is at least in qualitative agreement with the 
analysis of Ref.\ \cite{reynaud01}. 
Investigating the orbital symmetry breaking in LiNiO$_2$ in terms of 
complex orbitals is fascinating from the point of view of symmetry: one 
has only 2 orbitals and 2 different exchange couplings. Also in that case 
we found a coexistence of FM and AFM couplings which may explain the absence 
of magnetic long range order at low temperature in LiNiO$_2$. But the actual 
signs of exchange between different and equal orbitals are just reversed in 
comparison with the analysis of Ref.\ \cite{reynaud01}. 

We thank especially J.-L.\ Richard for important clarifications concerning 
the symmetry of the Hamiltonian. We thank also A.\ Stepanov and D. Khomskii 
for stimulating
discussions. We acknowledge D.\ Mertz for sharing the results of unpublished 
bandstructure calculations.


\begin{thebibliography}{30}
\bibitem{goodenough58}
J.B.\ Goodenough, D.G.\ Wickham, and W.J.\ Croft, J.\ Phys.\ Chem.\ Sol.\ 
{\bf 5}, 107 (1958).
\bibitem{yamaura96}
K.\ Yamaura, M.\ Takano, A.\ Hirano, R.\ Kanno, J.\ Solid State Chem.\ 
{\bf 127}, 109 (1996).
\bibitem{reynaud01}
F.\ Reynaud, D.\ Mertz, F.\ Celestini, J.-M.\ Debierre, A.M.\ Ghorayeb,
P.\ Simon, A.\ Stepanov, J.\ Voiron, and C.\ Delmas, Phys.\ Rev.\ Lett.\ 
{\bf 86}, 3638 (2001).
\bibitem{bongers66}
P.F.\ Bongers and U.\ Enz, Solid State Commun.\ {\bf 4}, 153 (1966).
\bibitem{chappel00}
E.\ Chappel, M.D.\ N\'u\~nez-Regueiro, F.\ Dupont, G.\ Chouteau, 
C.\ Darie, and A.\ Sulpice, Eur.\ Phys.\ J.\ B {\bf 17}, 609 (2000).
\bibitem{feiner97}
L.F.\ Feiner, A.M.\ Ole\'s, and J.\ Zaanen, Phys.\ Rev.\ Lett.\ 
{\bf 78}, 2799 (1997).
\bibitem{li98}
Y.Q.\ Li, M.\ Ma, D.N.\ Shi, and F.C.\ Zhang, Phys.\ Rev.\ Lett.\ 
{\bf 81}, 3527 (1998).
\bibitem{bossche00}
M.\ van der Bossche, P.\ Azaria, P.\ Lecheminant, and F.\ Mila, 
Eur.\ Phys.\ J.\ B {\bf 17}, 367 (2000).
\bibitem{mostovoy02}
M.V.\ Mostovoy and D.I.\ Khomskii, cond-mat/0201420 (2002).
\bibitem{GKA}
J. Goodenough, Phys.\ Rev.\  {\bf 100}, 564  (1955);
J. Kanamori, J. \ Phys. \ Chem. \ Solids {\bf 10}, 87  (1959);
P. W. Anderson, Phys.\ Rev.\  {\bf 115}, 2  (1959).
\bibitem{brink01}
J.\ van den Brink and D.\ Khomskii, Phys.\ Rev.\ B {\bf 63}, 1401416 (R) 
(2001).
\bibitem{khomskii}
 D.\ Khomskii Int. J. Mod. Phys. B15 ({\bf 19-20}),2865 (2001). 
\bibitem{maezono00}
R.\ Maezono and N.\ Nagaosa, Phys.\ Rev.\ B {\bf 62}, 11576 (2000).
\bibitem{fazekas} 
P. Fazekas Electron Correlation and Magnetism, World Scientific, (1999).
\bibitem{cyrot} M. Cyrot and C. Lyon-Caen, J.\ Phys {\bf 36},253, (1975).
\bibitem{dagotto96}
E.\ Dagotto, J.\ Riera, A.\ Sandvick, and A.\ Moreo, Phys.\ Rev.\ Lett.\ 
{\bf 76}, 1731 (1996).
\bibitem{mertz02}
D.\ Mertz, unpublished (2002).




\end{thebibliography}
\end{document}